# Longitudinal Speech Biomarkers for Automated Alzheimer's Detection

Jordi Laguarta[1] and Brian Subirana[1,2]*

[1] MIT AutoID Laboratory, Cambridge, MA, United States, [2] Faculty of Arts and Sciences, Harvard University, Cambridge, MA, United States

We introduce a novel audio processing architecture, the Open Voice Brain Model (OVBM), improving detection accuracy for Alzheimer's (AD) longitudinal discrimination from spontaneous speech. We also outline the OVBM design methodology leading us to such architecture, which in general can incorporate multimodal biomarkers and target simultaneously several diseases and other AI tasks. Key in our methodology is the use of multiple biomarkers complementing each other, and when two of them uniquely identify different subjects in a target disease we say they are orthogonal. We illustrate the OBVM design methodology by introducing sixteen biomarkers, three of which are orthogonal, demonstrating simultaneous above state-of-the-art discrimination for two apparently unrelated diseases such as AD and COVID-19. Depending on the context, throughout the paper we use OVBM indistinctly to refer to the specific architecture or to the broader design methodology. Inspired by research conducted at the MIT Center for Brain Minds and Machines (CBMM), OVBM combines biomarker implementations of the four modules of intelligence: The brain OS chunks and overlaps audio samples and aggregates biomarker features from the sensory stream and cognitive core creating a multi-modal graph neural network of symbolic compositional models for the target task. In this paper we apply the OVBM design methodology to the automated diagnostic of Alzheimer's Dementia (AD) patients, achieving above state-of-the-art accuracy of 93.8% using only raw audio, while extracting a personalized subject saliency map designed to longitudinally track relative disease progression using multiple biomarkers, 16 in the reported AD task. The ultimate aim is to help medical practice by detecting onset and treatment impact so that intervention options can be longitudinally tested. Using the OBVM design methodology, we introduce a novel lung and respiratory tract biomarker created using 200,000+ cough samples to pre-train a model discriminating cough cultural origin. Transfer Learning is subsequently used to incorporate features from this model into various other biomarker-based OVBM architectures. This biomarker yields consistent improvements in AD detection in all the starting OBVM biomarker architecture combinations we tried. This cough dataset sets a new benchmark as the largest audio health dataset with 30,000+ subjects participating in April 2020, demonstrating for the first time cough cultural bias.

# 1. INTRODUCTION

Since 2001, the overall mortality for Alzheimer's Dementia (AD) has been increasing year-on-year. Between 2000 and 2020 deaths resulting from stroke, HIV and heart disease decreased while reported deaths from AD increased by about 150% (Alzheimer's Association, 2020). Currently no treatments are available to cure AD, however, if detected early on, treatments may greatly slow and eventually possibly even halt further deterioration (Briggs et al., 2016).

Currently, methods for diagnosing AD often include neuroimaging such as MRI (Fuller et al., 2019), PET scans of the brain (Ding et al., 2019), or invasive lumbar puncture to test cerebrospinal fluid (Shaw et al., 2009). These diagnostics are far too expensive for large-scale testing and are usually used once family members or personal care detect late-stage symptoms, when the disease is too advanced for onset treatment. On top of the throughput limitations, recent studies on the success of the most widely used form of diagnostic, PET amyloid brain scans, have shown expert doctors in AD currently misdiagnose patients in about 83% of cases and change their management and treatment of patients nearly 70% of the time (James et al., 2020). This is mainly caused by the lack of longitudinal explainability of these scans. As a result it is hard to track effectiveness of treatments and even more to evaluate personalized treatments tailored to specific on-set populations of AD (Maclin et al., 2019). AI in general suffers from similar issues and operates a bit as a black-box, and does not offer explainable results linked to specific causes of each individual subject (Holzinger et al., 2019).

Based on the above findings, our research aims to find AD diagnostic methods achieving the following four warrants:

1. **Onset Detection:** detection needs to occur as soon as the first signs emerge, or sooner even if only probabilistic metrics can be provided. Preclinical AD diagnosis and subsequent treatment may offer the best chances at delaying the effects of dementia (Briggs et al., 2016). Therapeutic significance may require establishing subclassifications within AD (Briggs et al., 2016). Evidence that there are early signs of AD onset in the human body come in the form of recent research on blood plasma phosphorylated-tau isoforms diagnostic biomarkers demonstrating chemical traces of dementia, and of AD in particular, decades in advance of clinical diagnosis (Barthélemy et al., 2020; Palmqvist et al., 2020). These are encouraging findings, and hopefully there are also early onset signs in free-speech audio signals. In fact, preclinical AD is often linked to mood changes and in cognitively normal adults onset AD includes depression (Babulal et al., 2016), while apathy and anxiety have been linked to some cognitive decline (Bidzan and Bidzan, 2014). Both of these may be detectable in preclinical AD using existing sentiment analysis techniques (Zunic et al., 2020).
2. **Minimal Cost:** we need a method that has very little side effects, so that a person can perform the test periodically, and at very low variable costs to allow broad pre-screening possibilities. Our suggestion is to develop methods that can run on smart speakers and mobile phones (Subirana et al., 2017b) at essentially no cost while respecting user privacy (Subirana et al., 2020a). There is no medically approved system allowing preclinical AD diagnosis at scale. There are different approaches to measure AD disease onset and progression but all rely on expensive human assessments and/or medical procedures. We demonstrate our approach using only free speech but the approach can also include multi-modal data if available including MRI images (Altinkaya et al., 2020) and EEG recordings (Cassani et al., 2018).
3. **Longitudinal tracking:** the method should include some form of AD degree metric, especially to evaluate improvements resulting from medical interventions. The finer disease progression increments can be measured, the more useful they'll be. Ideally, adaptive clinical trials would be supported (Coffey and Kairalla, 2008).
4. **Explainability:** the results need to have some form of explainability, if possible including the ability to diagnose other types of dementia and health conditions. Most importantly, the approach needs to be approved for broad use by the medical community.

**TABLE 1** | A review of other AD diagnostic algorithms on the same dataset from Lyu (2018).

| References | Date | Accuracy(%) |
| --- | --- | --- |
| Syed et al. (2020) | 2020 | 85.4 |
| Haulcy and Glass (2021) | 2021 | 85.4 |
| Orimaye et al. (2014) | 2016 | 87.5 |
| Yuan et al. (2020) | 2020 | 89.6 |
| Karlekar et al. (2018) | 2018 | 91.1 |
| Laguarta and Subirana | 2021 | 93.8 |

*Our top performing model only uses audios while Orimaye et al. only used 35 patients hence risking high variance. Karlekar et al. only used transcripts. The rest used the transcripts from the ADreSS challenge Luz et al. (2020).*

Our approach is enabled by and improves upon advances in deep learning on acoustic signals to detect discriminating features between AD and non-AD subjects—it aims to address the warrants above, including explainability which has been challenging for previous approaches. While research in AD detection from speech has been ongoing for several years most approaches did not surpass the 90% detection mark as shown in **Table 1**. These approaches use black-box deep learning algorithms providing little to no explainability as to what led the model's decision, making it hard for clinicians to use and hence slowing adoption by the healthcare system. In Petti et al. (2020), review of the literature on AD speech detection, about two thirds of the papers reviewed use Neural Nets or Support Vector Machines, while the rest focus on Decision Trees and Naïve Bayes. Neural Nets seem to achieve the highest detection accuracy on average. Previous work, instead, has very little inspiration on the different stages of human intelligence and at most focuses solely on modeling a small part of the brain as shown in Nassif et al. (2019), de la Fuente Garcia et al. (2020), and Petti et al. (2020).

Combining independent biomarkers with recent advances in our understanding of the four modules of the human brain as researched at MIT's Center for Brain Minds and Machines (CBMM) (CBM, 2020), we introduce a novel multi-modal processing framework, the MIT CBMM Open Voice Brain Model (OVBM). The approach described in this paper aims to overcome limitations of previous approaches, firstly by training the model on large speech datasets and using transfer learning so that the accurate learned features improve AD detection accuracy even if the sample of AD patients is not large. Secondly, by providing an explainable output in the form of a saliency chart that may be used to track the evolution of AD biomarkers.

The use of independent biomarkers in the CBMM Open Voice Brain Model enables researching what is the value of each of them, simply by contrasting results with and without one of the biomarkers—we illustrate this point with a biomarker focused on cough discrimination (Subirana et al., 2020b) and one focused on wake words (Subirana, 2020). We feel this is an original contribution of our work grounded on the connection between respiratory conditions and Alzheimer's.

Furthermore, we also show that our framework lets apply the same biomarker models for audio detection of multiple diseases, and explore whether there may be common biomarkers between AD and other diseases. To that end, the OVBM framework we introduce may be extended to various other tasks such as speech segmentation and transcription. It has already proven to detect COVID-19 from a forced-cough recording with high sensitivity including 100% asymptomatic detection (Laguarta et al., 2020). Here we demonstrate it in the individualized and explainable diagnostic of Alzheimer's Dementia (AD) patients, where, as shown in **Table 1** we achieve above state-of-the-art accuracy of 93.8% (Pulido et al., 2020), and using only raw audio as input, while extracting for each subject a saliency map with the relative disease progression of 16 biomarkers. Even with expensive CT scans, to date experts can not create consistent biomarkers as described in James et al. (2020), Henriksen et al. (2014), and Morsy and Trippier (2019) even when including emotional biomarkers, unlike our approach which automatically develops them from free speech. Experts point at this lack of biomarkers as the reason why no new drug has been introduced in the last 16 years despite AD (Zetterberg, 2019) being the sixth leading cause of death in the United States (Alzheimer's Association, 2019), and one of the leading unavoidable causes for loss of healthy life.

We found that cough features, in particular, are very useful biomarker enablers as shown in several experiments reported in this paper and that the same biomarkers could be used for COVID-19 and AD detection. Our emphasis on detecting relevant biomarkers corresponding to the different stages of disease onset, led us to build ten sub-models using four datasets. To do so, over 200,000 cough samples were crowd sourced to pre-train a model discriminating English from Catalan coughs, and then transfer learning was leveraged to exploit resulting features by integrating it into an OVBM brain model, showing improvements in AD detection, no matter what transfer learning strategy was used. This COVID-19 cough dataset we created approved by MIT's IRB 2004000133 sets a new benchmark as the largest audio health dataset, with over 30,000 subjects participating in less than four weeks in April 2020.

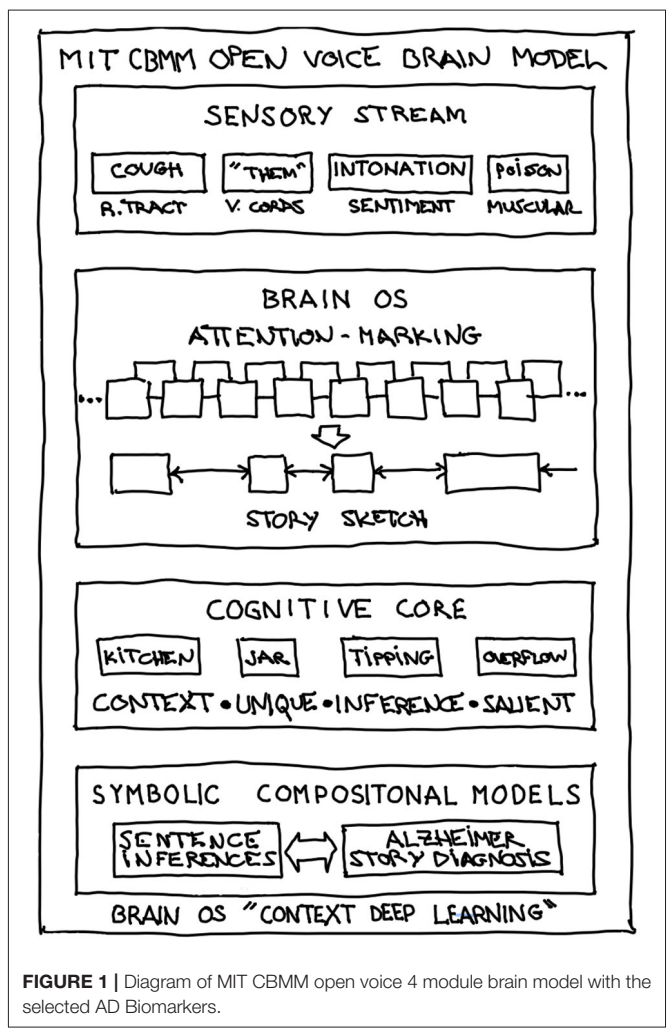

**FIGURE 1** | Diagram of MIT CBMM open voice 4 module brain model with the selected AD Biomarkers.

In the next section we present a literature review with evidence in favor of our choice of four biomarkers. In section 3, we present the different components of the Open Voice Brain Model AD detector, from sections 4 to 7 we introduce the 16 biomarkers with results and a novel personalized AD biomarker comparative saliency map. We conclude in section 8 with a brief summary and implications for future research.

## 2. LITERATURE REVIEW SUPPORTING OUR CHOICE OF FOUR SENSORY STREAM AUDIO BIOMARKERS: COUGH, WAKE WORD, SENTIMENT, AND MEMORY

Informed by a review of the literature, our choice of biomarkers is consistent with the vast literature resulting from AD research as we discuss next.

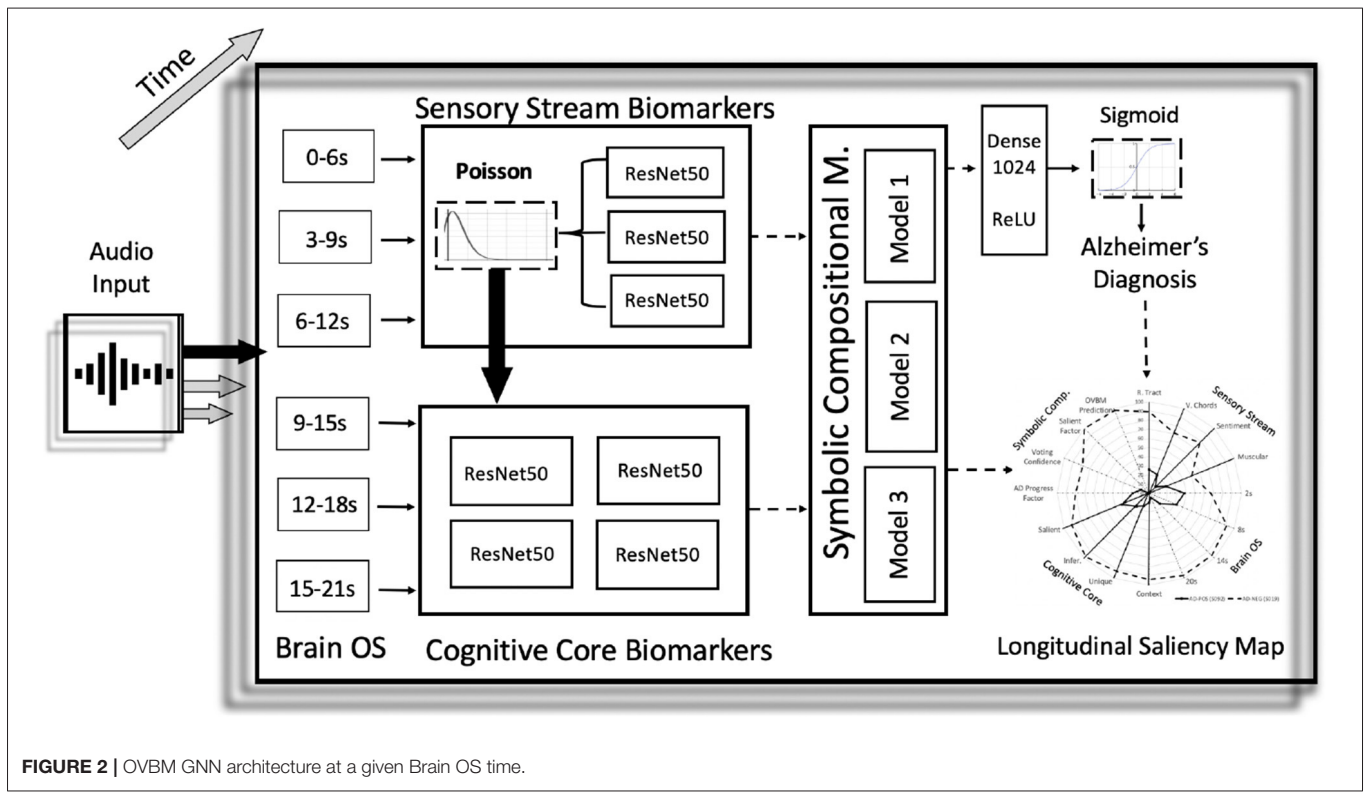

FIGURE 2 | OVBM GNN architecture at a given Brain OS time.

## 2.1. Mood Biomarkers

Preclinical AD is often linked to mood changes. In cognitively normal adults it include depression (Babulal et al., 2016), while apathy and anxiety have been linked to some cognitive decline (Bidzan and Bidzan, 2014). Sentiment biomarker. Clinical evidence supports the importance of sentiments in AD early-diagnosis (Costa et al., 2017; Galvin et al., 2020), and different clinical settings emphasize different sentiments, such as doubt, or frustration (Baldwin and Farias, 2009).

## 2.2. Memory Biomarkers

One of the main early-stage AD biomarkers is memory loss (Chertkow and Bub, 1990), which occurs both at a conceptual level as well as at a muscular level (Wirths and Bayer, 2008) and is different from memory forgetting in healthy humans (Cano-Cordoba et al., 2017; Subirana et al., 2017a). A prominent symptom of early stage AD is malfunctioning of different parts of memory depending on the particular patient (Small et al., 2000), possibly affecting one or more of its subcomponents including primary or working memory, remote memory, and semantic memory. The underlying causes of these memory symptoms may be linked to neuropathological changes, such as tangles and plaques, initially affecting selected areas of the brain like the hippocampi or the temporal and frontal lobes, and gradually expanding beyond these (Morris and Kopelman, 1986). Memory biomarker.

## 2.3. Respiratory Tract Biomarkers Cough and Wake Word

The human cough is already used to diagnose several diseases using audio recognition (Abeyratne et al., 2013; Pramono et al., 2016) as it provides information corresponding to biomarkers in the lungs and respiratory tract (Bennett et al., 2010). People with chronic lung disorders are more than twice as likely to have AD (Dodd, 2015), therefore we hypothesize features extracted from a cough classifier could be valuable for AD diagnosis.

There is an extensive cough-based diagnosis research of respiratory diseases but to our knowledge, no one had applied it to discriminate other, apparently unrelated, diseases like Alzheimer's. Our findings are consistent with the notion that AD patients cough differently and that cough-based features can help AD diagnosis; they are also consistent with the notion that cough features may help detect the onset of the disease. The lack of longitudinal datasets prevents us from exploring this point but do allow us to demonstrate the diagnostic power of cough-based features, to the point where without these features we would not have surpassed state-of-the-art performance.

The respiratory tract is often involved in the fatal outcome of AD. We introduce two biomarkers focused on the respiratory tract that may help discriminate between early and late stage AD. We have not found research indicating how early changes in the tract may be detected but given it's importance in the disease outcome it may be early on. This could also explain the success of many speech-based AD discrimination approaches— some of which have been applied to early stages of FTD. Significant research in AD such as Heckman et al. (2017)

**TABLE 2** | Impact of Poisson mask on AD performance.

| Model | W/o Poisson(%) | With Poisson(%) |
|---|---|---|
| Baseline | 65.6 | 68.8 |
| Cough | 75.0 | 75.0 |
| Intonation | 68.8 | 75.0 |
| Wake-Word "Them" | 75.0 | 78.1 |
| Multi-Modal | 90.6 | 93.8 |
| Avg improvement(%) | | 3.1 |

*Baseline is a ResNet50 trained on the AD task without transfer learning.*

**TABLE 3** | To illustrate the complementary nature of the biomarkers we show the unique AD patients detected by each individual biomarker model with only the final classification layer fine-tuned on the target disease, Alzheimer's and COVID-19 in this case.

| Biomarker | Model Name | Alzheimer's(%) | COVID-19(%) |
|---|---|---|---|
| Respiratory tract | Cough | 9 | 23 |
| Sentiment | Intonation | 19 | 8 |
| Vocal cords | WW "THEM" | 16 | 19 |
| R. Tract and sentiment | Cough and Tone. | 0 | 0 |
| R. Tract and vocal cords | Cough and WW | 6 | 1 |
| Sentiment and vocal cords | Tone. and WW | 3 | 0 |
| In all 3 | | 41 | 34 |
| In neither of the 3 | | 6 | 15 |

*Each transfer model detects unique patients reinforcing orthogonality of the biomarkers and hence the potential of combining new ones. Note how exactly the same biomarker models can detect Alzheimer's and COVID-19 subjects, showing the transferable nature for different diseases and how they behave "orthogonally" in both cases.*

has proven that the disease impacts motor neurons. In other diseases, like Parkinson's, where motor neurons are affected, vocal cords have proven to be one of the first muscles affected (Holmes et al., 2000).

Dementia in general has been linked to increased deaths from pneumonia (Wise, 2016; Manabe et al., 2019) and COVID-19 (Azarpazhooh et al., 2020; Hariyanto et al., 2020) possibly linked to specific gens (Kuo et al., 2020). COVID-19 deaths are more likely with Alzheimer's than with Parkinson's disease (Yu et al., 2020). This different respiratory response depending on the type of dementia suggests that related audio features, such as coughs, may be useful not only to discriminate dementia subjects from others but also to discriminate specific types of dementia.

We contend there is correlation, instead of causality, between our two respiratory track biomarkers and Alzheimer's but further elucidation to this extent is necessary as there is in many other areas with AD and more broadly in science in general (Pearl and Mackenzie, 2018). Some causality link may exist due to the simultaneous role of substance P in Alzheimer's (Severini et al., 2016) and in cough (Sekizawa et al., 1996). The existence of spontaneous cough *per se* may not be enough to predict onset risk but in combination with other health parameters may contribute to an accurate risk predictor (Song et al., 2011). Our biomarker suggestion is based on "forced coughs" which, to our knowledge, has not been studied in connection with Alzheimer's. We feel it may be an early indication of future respiratory tract conditions that will show in the form of spontaneous coughs. In patients with late-onset Alzheimer's Disease (LOAD) a unique delayed cough response has been reported in COVID-19 infected subjects (Isaia et al., 2020; Guinjoan, 2021). Dysphagia and aspiration pneumonia continue to be the two most serious conditions in late stage AD with the latter being the most common cause of death of AD patients (Kalia, 2003), suggesting substance P induced early signs in the respiratory tract may already be present in forced coughs, perhaps even unavoidably.

What seems unquestionable is the connection between speech and orofacial apraxia and Alzheimer's, and it has been suggested that it, alone, can be a good metric for longitudinal assessment (Cera et al., 2013). Various forms of apraxia have been linked to AD progression in different parts of the brain (Giannakopoulos et al., 1998). Nevertheless, given the difficulty in estimating speech and orofacial apraxia these figures are not part of common Clinical Dementia Rating scales (Folstein et al., 1975; Hughes et al., 1982; Clark and Ewbank, 1996; Lambon Ralph et al., 2003). However, all these studies reveal difficulties in an objective, accurate, and personalized scale that can track each patient independently from the others (Olde Rikkert et al., 2011). The lack of metrics also spans other related indicators such as quality of life estimations (Bowling et al., 2015). There are no reliable biomarkers for other neurogenerative disorders either (Johnen and Bertoux, 2019).

Recent research has demonstrated that apraxia screening can also predict dementia disease progression (Pawlowski et al., 2019), especially as a way to predict AD in early stage FTD subjects, a population that we are particularly interested in targeting with our biomarkers. For the Behavioral Variant of Fronto Temporal Dementia (bvFTD), in patients under 65 the second most common cognitive disorder caused by neurodegeneration, little tonal modulation and buccofacial apraxia, are targeted by our biomarkers and are established diagnostic domains (Johnen and Bertoux, 2019). We hope that our research can help establish reliable biomarkers for disease progression that can also distinguish at onset between the different possible diagnostics. The exact connection between buccofacial apraxia and dementia has not been as well-documented as that of other forms of apraxia. Recent results show that there buccofacial apraxia may be present in up to fifty percent of dementia patients with no association to oropharyngeal dysphagia (Michel et al., 2020). Oropharyngeal dysphagia, on the other hand, has been linked to dementia, in some studies in over fifty percent of the cases, appearing, in particular, in late stages of FTD and in early stages of AD (Alagiakrishnan et al., 2013).

According to the NIH's National Institute of Neurological Disorders and Stroke information page on apraxia[1], the most common form of apraxia is orofacial apraxia which causes the inability to carry out facial movements on request such as coughing. Cough reflex sensitivity and urge-to-cough deterioration has been shown to help distinguish AD from dementia with Lewy Bodies and control groups (Ebihara et al., 2020). The impairment of cough in the elderly is linked to dementia (Won et al., 2018).

---

[1] https://www.ninds.nih.gov/disorders/all-disorders/apraxia-information-page.

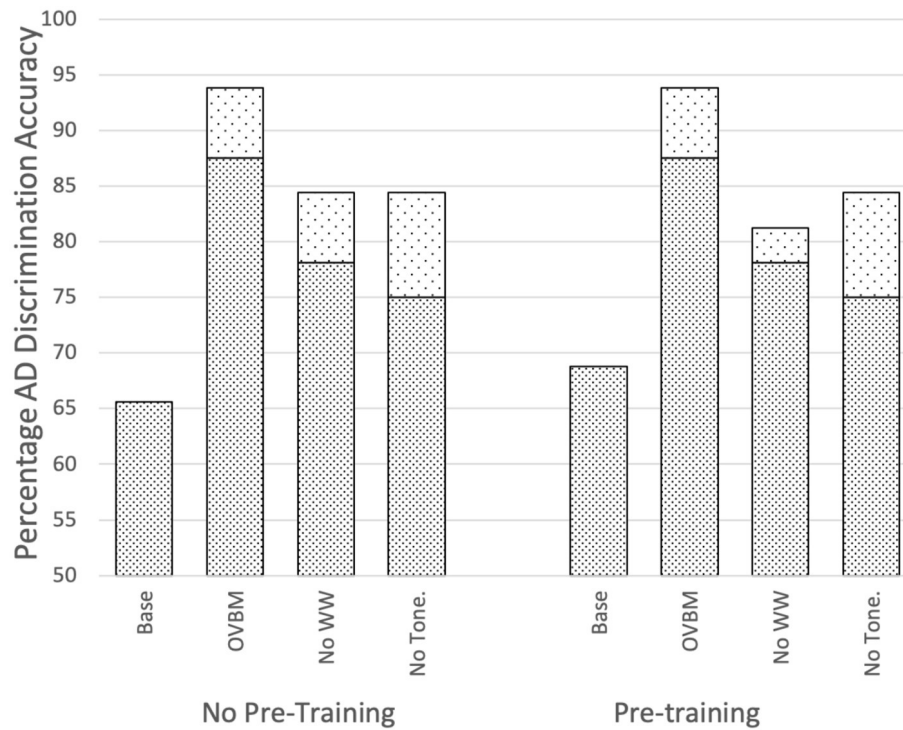

FIGURE 3 | Impact of sensory stream biomarkers on OVBM performance by removing transferred knowledge one at a time. Top dotted sections of bars indicate there is always performance gain from the cough biomarker. Baselines are the OVBM trained on AD without any transfer learning. In the other bars, a biomarker is removed and replaced with an AD pre-trained ResNet50, hence removing the transferred knowledge but conserving computational power, showing complementarities since all are needed for maximum results.

## 3. OVERVIEW OF THE MIT OPEN VOICE BRAIN MODEL (OVBM) FRAMEWORK

The OVBM architecture shown in **Figure 1** frames a four-unit system to test biomarker combinations and provides the basis for an explainable diagnostic framework for a target task such as AD discrimination. The Sensory Stream is responsible for pre-training models on large speech datasets to extract features of individual physical biomarkers. The Brain OS splits audio into overlapping chunks and leverages transfer learning strategies to fine-tune the biomarker models to the smaller target dataset. For longitudinal diagnosis, it includes a round-robin five stage graph neural network that marks salient events in continuous speech. The Cognitive Core incorporates medical knowledge specific to the target task to train cognitive biomarker feature extractors. The Symbolic Compositional Models unit combines fine-tuned biomarker models into a graph neural network. Its predictions on individual audio chunks are fed into an aggregating engine subsequently reaching a final diagnostic plus a patient saliency map. To enable doctors to gain insight into the specific condition of a given patient, one of the novelties of our approach is that the outputs at each unique module are extracted to create a visualization in the form of a health diagnostic saliency map showing the impact of the selected biomarkers. This saliency map may be used to longitudinally track and visualize disease progression.

### 3.1. OVBM Applied to AD Detection

Next, we review each of the four OVBM modules in the context of AD, introducing 16 biomarkers and gradually explaining the partial GNN architecture shown in **Figure 2**. To be able to compare models, our baselines and 8 of the biomarkers are based on the ResNet50 CNN due to its state-of-the-art performance on medical speech recognition tasks (Ghoniem, 2019). All audio samples are processed with the MFCC package published by Lyons et al. (2020), and padded accordingly. We operate on Mel Frequency Cepstral Coefficients (MFCC), instead of spectrograms (Lee et al., 2009), because of its resemblance to how the human cochlea captures sound (Krijnders and t Holt, 2017). All audio data uses the same MFCC parameters (Window Length: 20 ms, Window Step: 10 ms, Cepstrum Dimension: 200, Number of Filters: 200, FFT Size: 2,048, Sample rate: 16,000). All datasets follow a 70/30 train-test split and models are trained with an Adam optimizer (Kingma and Ba, 2014).

The dataset from DementiaBank, ADrESS (Luz et al., 2020), is used for training the OVBM framework and fine-tuning all biomarker models on AD detection. This dataset is the largest publicly available, consisting of subject recordings in full enhanced audio and short normalized sub-chunks, along with the recording transcriptions from 78 AD and 78 non-AD patients. The patient age and gender distribution is balanced and equal for AD and non-AD patients. For the approach of this study focusing purely on audio processing we only use the full enhanced audio and patient metadata, excluding transcripts from any processing. It is worth noting this given the poor audio quality of some of the recordings.

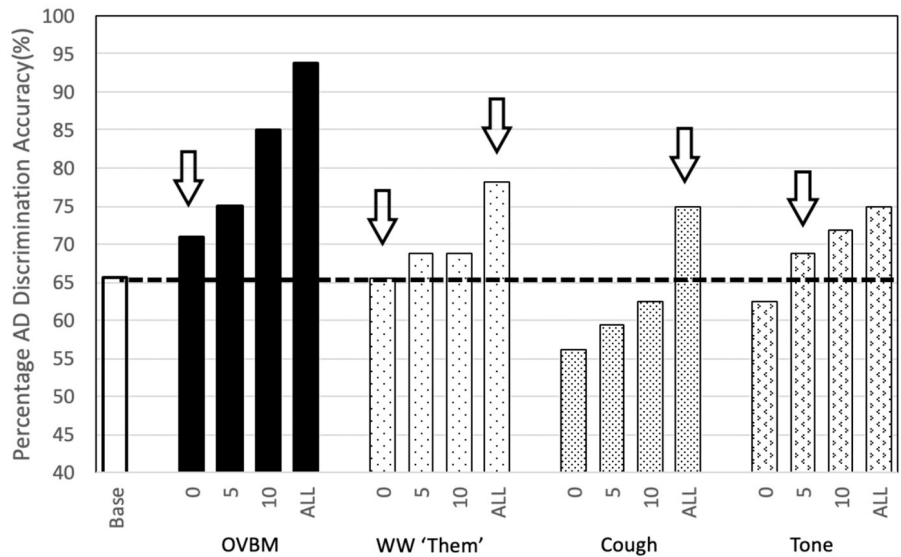

**FIGURE 4** | Sensory Stream Saliency Bar Chart: To illustrate the potential of our approach we show the strength of the simplest transfer models we tried. The numbers 0-5-10-ALL on the x-axis labels refer to the number of convolution layers trained after transfer learning in addition to the final dense layer. We find the most surprising, perhaps, is that the simple wakeword model to find the word "Them" is as powerful as the baseline. If we let the model fine-tune the last few (0-5-10) layers then it goes well beyond it. Our novel Cough database, inspired in the effect of AD in the respiratory tract also shows surprising results, even without any adaptation at all. If we let fine-tuning of the whole model, it's validation accuracy improves ≈10% points with respect to the baseline. Baseline is the same OVBM architecture trained on AD without any transfer learning of features.

## 4. OVBM AD SENSORY STREAM BIOMARKERS

We have selected four biomarkers inspired by previous medical community choices (Chertkow and Bub, 1990; Wirths and Bayer, 2008; Dodd, 2015; Heckman et al., 2017; Galvin et al., 2020), as reviewed next.

### 4.1. Biomarker 1 (Muscular Degradation)

We follow memory decay models from Subirana et al. (2017a) and Cano-Cordoba et al. (2017) to capture this muscular metric by degrading input signals for all train and test sets with the **Poisson** mask shown in Equation (1), a commonly occurring distribution in nature (Reed and Hughes, 2002). We use as a Possion function a mask with input MFCC image = $I_x$, output mask = $M(I_X)$, $\lambda = 1$, and k = each value in $I_x$:

$$M(I_x) = Pr(\lambda)I_x \qquad (1)$$

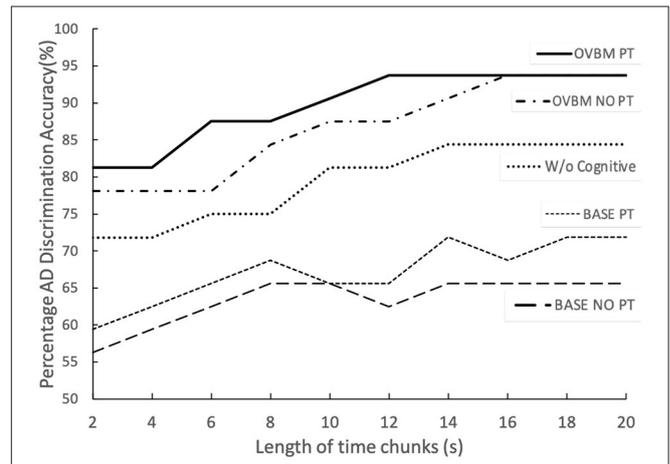

**FIGURE 5** | The two top lines illustrate the full OVBM performance, with its biomarker feature models, as a function of chunk size. PT refers to individually fine-tuning each biomarker model for AD before re-training the whole OVBM. The middle line shows the OVBM without the cognitive core, illustrating how it boosts performance by about 10% across the board. Baseline PT is the OVBM architecture with each ResNet50 inside individually trained on AD before retraining them together in the OVBM architecture.

$$Pr(X = k) = \frac{\lambda^k e^{-k}}{k!}$$

As shown in **Table 2**, this Poisson biomarker brings a unique improvement to each model except for Cough, consistent with both inherently capturing similar features containing muscular degradation.

## 4.2. Biomarker 2 (Vocal Cords)

We have developed a vocal cord biomarker to incorporate in OBVM architectures. We trained a Wake Word (WW) model to learn vocal cord features on LibriSpeech—an audiobook dataset with ≈1,000 h of speech from Panayotov et al. (2015) by creating a balanced sample set of 2 s audio chunks, half containing the word "Them" and half without. A ResNet50 (He et al., 2016) is trained for binary classification of **"Them"** on 3 s audio chunks(lr:0.001, val_acc: 89%).

Illustrated in **Table 3** and **Figure 4**, this vocal cords model proves its contribution of unique features, which without fine-tuning to the AD task performs as well as the baseline ResNet50 fully trained on AD, and significantly beats it when fully fine-tuned.

## 4.3. Biomarker 3 (Sentiment)

We train a Sentiment Speech classifier model to learn **intonation** features on RAVDESS—an emotional speech dataset by Livingstone and Russo (2018) of actors speaking in eight different emotional states. A ResNet50 (He et al., 2016) is trained on categorical classification of eight corresponding intonations such as calm, happy, or disgust(lr: 0.0001, val_acc on 8 classes: 71%).

As illustrated by **Table 3** and **Figure 4**, this biomarker captures unique features for AD detection, and when only fine-tuning its final five layers outperforms a fully trained ResNet50 on AD detection.

## 4.4. Biomarker 4 (Lungs and Respiratory Tract)

We use the **cough** dataset collected through MIT Open Voice for COVID-19 detection (Subirana et al., 2020b), strip all but the spoken language of the person coughing (English, Catalan), and split audios into 6 s chunks. A ResNet50 (He et al., 2016) is trained on binary classification (Input: MFCC 6s Audio Chunks (1 cough)—Output: English/Catalan, lr: 0.0001, val_acc: 86%).

**Figure 4** and **Table 3**, justify the features extracted by this cough model as valuable for the task of AD detection by capturing a unique set of samples and improving performance. Further, **Figure 3** validates its impact on various OBVM architectures, including the top performing multi-modal model, justifying the relevance of this novel biomarker.

## 5. OVBM BRAIN OS BIOMARKERS

The Brain OS is responsible for capturing learned features from the individual biomarker models in the Sensory Stream and Cognitive Core, and for integrating them into an OVBM architecture, with the aim of training the ensemble for a target task, in this case AD detection.

To make the most out of the short patient recordings, we split each patient recording into overlapping audio chunks (0–4, 2–6, 4–8 s). Once the best pre-trained biomarker models in the sensory stream and cognitive modules are selected, we first concatenate them together and then pass their outputs through a 1,024 neuron deeply connected neural network layer with ReLU activation. We also incorporate at this point metadata such as gender. We test three Brain OS transfer learning strategies: (1) CNNs are used as fixed feature extractors without any fine-tuning; (2) CNNs are fine-tuned by training all layers; (3) only the final layers of the CNN are fine-tuned.

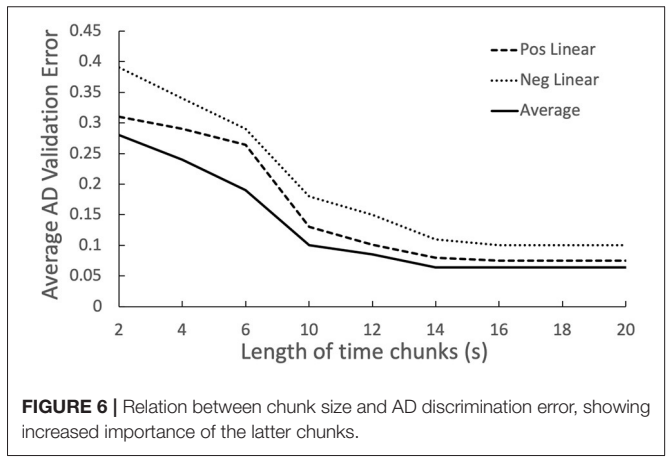

**FIGURE 6** | Relation between chunk size and AD discrimination error, showing increased importance of the latter chunks.

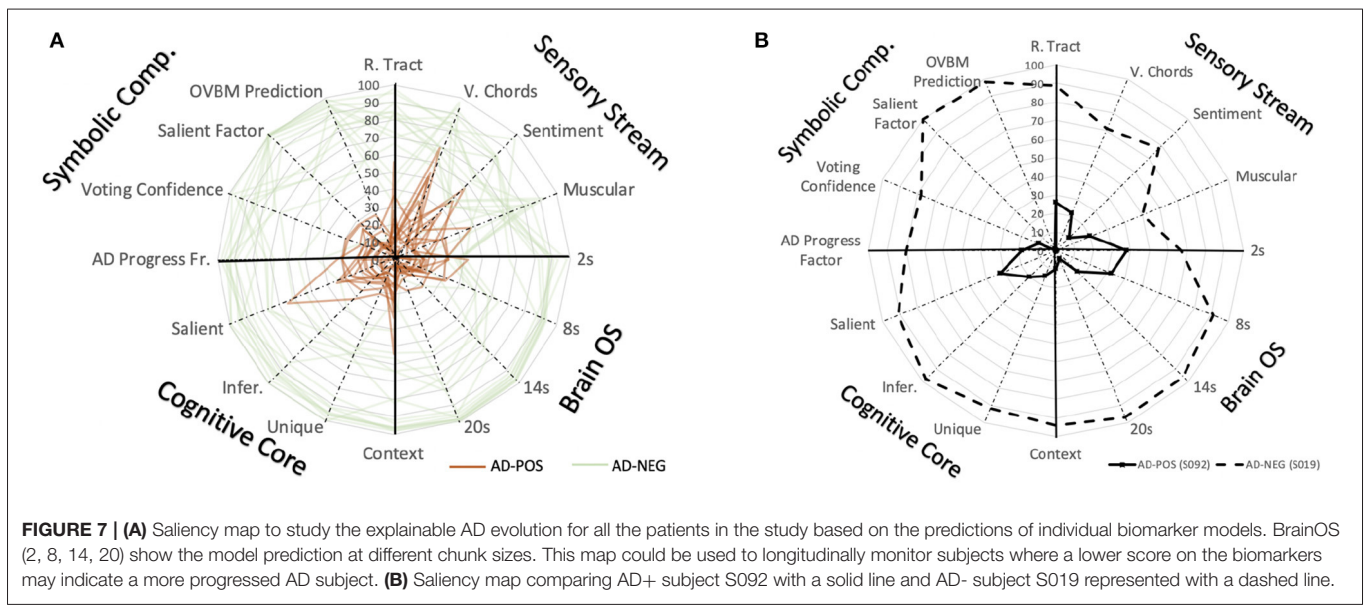

**FIGURE 7** | **(A)** Saliency map to study the explainable AD evolution for all the patients in the study based on the predictions of individual biomarker models. BrainOS (2, 8, 14, 20) show the model prediction at different chunk sizes. This map could be used to longitudinally monitor subjects where a lower score on the biomarkers may indicate a more progressed AD subject. **(B)** Saliency map comparing AD+ subject S092 with a solid line and AD- subject S019 represented with a dashed line.

From **Figure 5**, it is evident AD detection improves as chunk length increases consistent with the fact that attention-marking has more per-chunk information to formulate a better AD prediction. From this attention-marking index (quantity of information required in a chunk for a confident diagnosis) we select chunk sizes **2**, **8**, **14**, and **20 s**, shown in **Figure 7**, as the Brain OS biomarkers, establishing individual AD progression. In terms of transfer learning strategies, **Figure 4** shows that fine-tuning all layers always leads to better results, however for most models almost no fine-tuning is required to beat the baseline.

## 6. OVBM COGNITIVE CORE BIOMARKERS

Neuropsychological tests are a common screening tool for AD (Baldwin and Farias, 2009). These tests, among others, evaluate a patient's ability to remember uncommon words, contextualize, infer actions, and detect saliency (Baldwin and Farias, 2009; Costa et al., 2017). In the case of this AD dataset, all patients are asked to describe the Cookie Theft picture created by Goodglass et al. (1983), where a set of words such as "kitchen" (**context**), "tipping" (**unique**), "jar" (**inferred**), and "overflow" (**salient**), are used to capture four cognitive biomarkers. To keep the richness of speech, we train four wake word models from LibriSpeech (Panayotov et al., 2015) with ResNet50s following the same approach as Biomarker 2. The four chosen cognitive biomarkers aim to detect patients' ability on: context, uniqueness, inference, and saliency.

We could show the same saliency bar chart in **Figure 4** and a uniqueness table such as **Table 3** to illustrate the impact of each cognitive biomarker. Instead in **Figure 5**, we show the impact of removing the cognitive core on the top OVBM performance which drops ≈10%, validating the relevance of the cognitive core biomarkers.

## 7. OVBM SYMBOLIC COMP. M. BIOMARKERS

This module fine-tunes previous modules' outputs into a graph neural network. Predictions on individual audio chunks for one subject are aggregated and fed into competing models to reach a final diagnostic. We tested the model with various BERT configurations and found no improvement in detection accuracy. In the AD implementation, given we had at most 39 overlapping chunks, three simple aggregation metrics are compared: averaging, linear positive (more weight given to later chunks), and linear negative (more weight given to earlier chunks).

In **Figure 6**, averaging proves to be the most effective, while positive linear over performing the negative linear indicates the latter audio chunks are more informative than front ones. **Figure 7** includes four biomarkers derived from combining chunk predictions from biomarker models of the three other modules (Cummings, 2019). With more data and longitudinal recordings, the OVBM GNN may incorporate other biomarkers.

## 8. DISCUSSION

We conclude by providing a few insights further supporting our OVBM brain-inspired model for audio health diagnostics as presented above. We have proven the success of the OVBM framework, setting the new benchmark for state-of-the-art accuracy of AD classification, despite only incorporating audio signals—one that can incorporate GNNs (Wu et al., 2020). Future work may improve this benchmark by also incorporating into OVBM longitudinal GNN's natural language biomarkers using NLP classifiers or multi-modal graph neural networks incorporating non-audio diagnostic tools (Parisot et al., 2018).

One of the most promising insights of all is the discovery of cough as a new biomarker (**Figure 3**), one that improves any of the intermediate models tested and that validates OVBM as a framework on which medical experts can hypothesize and test out existing and novel biomarkers. We are the first to report that cough biomarkers have information related to gender and culture, and are also the first to demonstrate how they improve simultaneous AD classification as illustrated in the saliency charts (**Figure 4**) as well as that of other apparently unrelated conditions.

Another promising finding is the model's explainability, introducing the biomarker AD saliency map tool, offering novel methods to evaluate patients longitudinally on a set of physical and neuropsychological biomarkers as shown in **Figure 7**. In future research, longitudinal data may be collected to properly test the onset potential of OVBM GNN discrimination in continuous speech. We hope our approach brings the AI health diagnostic experts closer to the medical community and accelerates research for treatments by providing longitudinal and explainable tracking metrics that can help succeed adaptive clinical trials of urgently needed innovative interventions.

## DATA AVAILABILITY STATEMENT

The data analyzed in this study is subject to the following licenses/restrictions: in order to gain access to the datasets used in the paper, researchers must become a member of DementiaBank. Requests to access these datasets should be directed to https://dementia.talkbank.org/.

## AUTHOR CONTRIBUTIONS

JL wrote the code. BS designed the longitudinal biomarker Open Voice Brian Model (OVBM) and the saliency map. All authors contributed to the analysis of the results and the article.

**Conflict of Interest:** The authors declare that the research was conducted in the absence of any commercial or financial relationships that could be construed as a potential conflict of interest.